\documentclass[10pt,twocolumn,a4paper]{article}

\usepackage[left=20mm, right=15mm, top=15mm, bottom=15mm]{geometry}

\usepackage{subfig}
\usepackage{flushend}
\usepackage{graphics}
\usepackage{amsmath}
\usepackage{graphicx}
\usepackage{epstopdf}
\usepackage{float}
\usepackage[english]{babel}
\begin{document}

\title{\huge \textbf{Analytical Computation of the Perihelion Precession in General Relativity via the Homotopy Perturbation Method}}

\date{}

\twocolumn[
\begin{@twocolumnfalse}
\maketitle

\author{\textbf{V.~K.~ Shchigolev}$^{1,*}$\\\\
\footnotesize $^{1}${Department of Theoretical Physics, Ulyanovsk State University, 42 L. Tolstoy Str., Ulyanovsk 432000, Russia}\\

\footnotesize $^{*}$Corresponding Author: vkshch@yahoo.com}\\\\\\

\end{@twocolumnfalse}
]

\noindent \textbf{\large{Abstract}} \hspace{2pt} We propose a new approach in studying  the
planetary orbits and the perihelion precession in General Relativity by means of the Homotopy Perturbation
Method (HPM).For this purpose, we give a brief review of the nonlinear geodesic equations in the spherical
symmetry spacetime which are to be studied in our work. On the basis of the main idea of HPM,
we construct the appropriate homotopy what leads to the problem of solving the set of linear equations.
First of all, we consider the simple example of the Schwarzschild metric  for which the approximate geodesics solutions
are known, in order to compare the HPM solution for orbits with those obtained earlier. Moreover,
we obtain an approximate HPM  solution for the Reissner-Nordstorm spacetime of a charged star.
\\

\noindent \textbf{\large{Keywords}} \hspace{2pt} Planetary orbits, General Relativity, Perihelion precession, Homotopy Perturbation Method.\\
\noindent\hrulefill

\section{\Large{Introduction}}
\quad

The advance of the perihelion in the orbit of Mercury is a relativistic effect \cite{Weinberg}.
Together with the observation of the deflection of light, this result offers unique possibilities
for testing General Relativity (GR) and exploring the limits of alternative theories of gravitation.

The calculations of the perihelion precession in General Relativity and some modified theories of
gravity, were recently considered in \cite{Magnan}-\cite{Pejic}. For example,  a way to obtain
information about higher dimensions from observations by studying a brane based spherically symmetric
solution is considered for the classic tests of General Relativity in \cite{Cuzinatto}.
The analytical computation of the Mercury perihelion precession in the frame of relativistic gravitational
law and comparison with general relativity is presented in \cite{Fokas}.

In \cite{Ruggiero},  the perturbations determined by a generic alternative theory of gravity to the GR
solution describing the gravitational field around a central mass are worked out.
In \cite{Saridakis}, the authors used recent observations from solar system orbital motions in
order to constrain $f(T)$ gravity. In particular,  the spherical solutions of the theory are used
to describe the Sun's gravitational field  and advances of planetary perihelia in order to obtain
upper bounds on the allowed $f(T)$ corrections.

It is well known that the geodesics equations in RG are nonlinear, and therefore cannot in general
be solved exactly. For instance, the geodesic equations resulting from the Schwarzschild gravitational
metric element are solved exactly by the Weierstra$\ss$ Jacobi modular form \cite{Kraniotis}. Mostly,
the perihelion precession of planetary orbits based on Einstein's equations had been calculated in different
approximations for a general spherically symmetric line element.

The idea of the Homotopy Perturbation Method which is a semi-analytical method was first proposed by
Dr. Ji-Huan He \cite{He}-\cite{He3} for solving differential and integral equations. Later, the method is applied
to solve the non-linear and non-homogeneous partial differential equations \cite{He4}.

The HPM has a significant advantage providing an analytical approximate solution to a wide range of
nonlinear problems of the fundamental and applied sciences \cite{He6}, \cite{Cveticanin}. Sometimes, this
method  allows to find  even an exact solution with the help of a few iteration \cite{Nourazar}.
This method and a wide spectrum of its application have been extensively developed for several years  by
numerous authors (see \cite{He7} and references therein).

Recently  there were studies in which this method was used for analytical calculations in the field of
cosmology and astrophysics (see, e.g. \cite{Shchigolev1}, \cite{Rahaman}).
Our aim is to give one more application of the method to the problem of planetary motion in the spherically
symmetric gravitational field in General Relativity.

\section{\Large{Geodesics in the Spherical Symmetry Spacetime}}
\quad

As well known \cite{Weinberg,Hu}, the line element of the 4-dimensional general spherically symmetric
stationary spacetime can be written as
\begin{equation}
ds^2=-f(r)dt^2+\frac{dr^2}{h(r)}+r^2(d\theta^2+\text{sin}^2\theta d \varphi^2), \label{1}
\end{equation}
and the perihelion precession is usually treated as the time-like  geodesic in spacetime. Let us consider
the geodesics $\gamma(\tau)$ in the above spherically symmetric spacetime. We set the geodesic $\gamma(\tau)$
expressed in the spherical coordinates $x^{\mu}=(t,r,\theta,\varphi)$ as $x^{\mu}(\tau)$, which are satisfied
$$
\frac{d^2x^{\mu}}{d\tau^2}+\Gamma^{\mu}_{\nu\sigma}\frac{dx^{\nu}}{d\tau}\frac{dx^{\sigma}}{d\tau}=0.
$$
The geodesic $\gamma(\tau)$ can be obtained by solving the above equation. However, taking into account  the
symmetry of spacetime (\ref{1}), one could use the following  simple way to obtain the geodesic $\gamma(\tau)$.

First, we can find that one component of the geodesic $\gamma(\tau)$ can always be chosen as $\theta (\tau)= \pi/2$,
which means that the geodesic can always be chosen to lay in the equatorial plane of the spherically symmetric spacetime.
Thus,
$t=t(\tau),~r=r(\tau),~\theta=\pi/2,~\varphi=\varphi(\tau)$.
Let us denote the tangent vector of geodesic $\gamma(\tau)$ as $U^{\mu}\equiv d x^{\mu}/d \tau$. For the time-like
geodesic, we chose $\tau$ to be the proper time . Hence, from (\ref{1})  we can obtain
\begin{equation}
\label{2}
f(r)\Big(\frac{dt}{d\tau}\Big)^2
-h^{-1}(r)\Big(\frac{dr}{d\tau}\Big)^2-r^2\Big(\frac{d\varphi}{d\tau}\Big)^2 = -1,
\end{equation}
where we have used $\theta=\pi/2$.

Second, it could be noted that $\xi^a =(\partial/\partial t)^a$ and $\psi^a=(\partial/\partial \varphi)^a$ are
two Killing vectors in the spherically symmetric spacetime (\ref{1}). Therefore, there are two conserved quantities
along the geodesic $\gamma(\tau)$, the total energy
\begin{equation}
E=-g_{ab} \xi^a U^b=f(r)\frac{dt}{d\tau}, \label{3}
\end{equation}
and the angular momentum per unit mass
\begin{equation}
L=g_{ab}\psi^a U^b=r^2\frac{d\varphi}{d\tau}. \label{4}
\end{equation}
After inserting (\ref{3}) and (\ref{4}) into (\ref{2}), one could obtain
\begin{equation}
\Big(\frac{dr}{d\tau}\Big)^2=\frac{h(r)}{f(r)}E^2-h(r)\Big(1+\frac{L^2}{r^2}\Big).
\label{5}
\end{equation}
This equation contains only one function $r(\tau)$, and it could be solved in principle. Then, after inserting the
solved $r(\tau)$ into (\ref{3}) and (\ref{4}), the rest components $t(\tau)$ and $\varphi(\tau)$ of geodesic could
be finally obtained.

However, it should be pointed that perihelion precession is usually related to the orbit of geodesic, i.e. $r(\varphi)$.
Therefore, it is convenient to rewrite the equation (\ref{5}) with the help of (\ref{4}) as
\begin{equation}
\Big(\frac{dr}{d\varphi}\Big)^2\Big(\frac{L}{r^2}\Big)^2=\frac{h(r)}{f(r)}E^2-h(r)\Big(1+\frac{L^2}{r^2}\Big).
\label{6}
\end{equation}
It has been found that the coordinate $u\equiv1/r$ is more convenient than $r$ to derive the perihelion precession.
Thus, the main equation investigated in our paper could be simply obtained from equation (\ref{6}) by converting $r$ into $u$
\begin{equation}
\Big(\frac{d u}{d\varphi}\Big)^2=\frac{h(u)}{f(u)}\Big(\frac{E}{L}\Big)^2-h(u)\Big(\frac{1}{L^2}+u^2\Big).
\label{7}
\end{equation}
Finally, differentiating the equation (\ref{7}) with respect to $\varphi$, we get the second-order geodesic equation
in the following form
\begin{equation}
\frac{d^2 u}{d\varphi^2}=\frac{E^2}{2 L^2}\frac{d }{d u}\left[\frac{h(u)}{f(u)}\right]-h(u)u-\frac{1}{2}
\Big(\frac{1}{L^2}+u^2\Big)\frac{d h(u)}{d u}. \label{8}
\end{equation}

\section{Homotopy Perturbation Method}
\quad
To illustrate the basic ideas of HPM \cite{He} for solving nonlinear differential equations, let us consider the
following nonlinear differential equation:
\begin{equation} \label{9}
A(u)=g(r), \,\,\, r \in \it{\Omega},
\end{equation}
with the boundary conditions $B(u, \partial u/\partial n)= 0;\, r \in \it{\Gamma}$,
where $A$ is a general differential operator, $B$ is a boundary operator, $g(r)$ is a known analytic function,
$\it{\Gamma}$ is the boundary of the domain $\it{\Omega}$. Suppose the operator $A$ can be divided into two parts:
$M$ and $N$. Therefore, (\ref{9}) can be rewritten as follows:
\begin{equation} \label{10}
M(u) +  N(u)= g(r).
\end{equation}
The homotopy $v(r, p): {\it{\Omega}}\times [0,1] \to {\it I\!\!R} $ constructed as follows \cite{He2}
\begin{equation} \label{11}
H(v, p)=(1- p) [M(v)- M(y_0)]+p\,[A(v)- g(r)] = 0,
\end{equation}
where $r \in {\it{\Omega}}$  and $p \in [0, 1]$ is an imbedding parameter, and $y_0$ is an initial approximation
of (\ref{9}). Hence, one can see that
\begin{equation} \label{12}
H(v, 0)= M(v)- M(y_0)=0,
H(v, 1)=A(v)- g(r) = 0,
\end{equation}
and changing the variation of $p$ from $0$ to $1$ is the same as changing $H(v, p)$ from $M(v)- M(y_0)$ to $A(v)- g(r)$,
which are called homotopic. In topology,
this is called deformation. Due to the fact that $0 \leq p \leq 1$ can be
considered as a small parameter, by applying the perturbation procedure, one can assume that the solution of (\ref{11})
can be expressed as a series in $p$, as follows:
\begin{equation}\label{13}
v=v_0+p\, v_1+p^2 v_2 + p^3 v_3 + ...\, .
\end{equation}
When we put $p \to 1$, then equation (\ref{11}) corresponds to (\ref{10}), and (\ref{13}) becomes the approximate
solution of (\ref{10}), that is $u(x)= \lim_{p \to 1} v =v_0+v_1+ v_2 +  v_3 + ...$.

This series is convergent for most cases. However, the convergent rate depends upon the nonlinear operator $A(v)$.
Sometimes, even the first approximation is sufficient to obtain the exact solution \cite{He}. As it is emphasized
in \cite{He2} and \cite{Cveticanin}, the second derivative of $N(v)$ with respect to $v$ must be small, because the
parameter $p$ may be relatively large, i.e. $p \to 1$, and the norm of $L^{-1} \partial N/\partial v$ must be smaller
than one, in order that the series converges.

\section{Orbits and Perihelion Precession via HPM}
\quad
First, we consider the simplest case of the metric (\ref{1}), namely, the Schwarzschild metric describing the
gravitational field of an uncharged non-rotating star.
For the Schwarzschild solution, the two functions are $f(r)=h(r)=1-2M/r$, or
\begin{equation}
f(u)=h(u)=1-2M u. \label{14}
\end{equation}
where $M$ is a mass of the star.  Therefore, Eq.(\ref{8}) for the time-like geodesic can be
\begin{equation}
\frac{d^2 u}{d\varphi^2}+u=\frac{M}{L^2}+3\,u^2\,M.
\label{15}
\end{equation}
Compared with the case in Newton's gravity
$$\frac{d^2 u}{d\varphi^2}+u=\frac{M}{L^2},$$
the term $3 u^2 M$ comes from the correction of general relativity. Note that, the analytical solution of (\ref{15})
is absent like the Schwarzschild case. However, there is an approximation solution of (\ref{15})
\begin{eqnarray}
u(\varphi)&=&\frac{M}{L^2}(1+e
\cos\varphi)\\ \nonumber
&+&\frac{M^3}{L^4}\Big(3+2e^2+3e\varphi \sin\varphi-e^2 \cos
^2\varphi\Big),
\label{16}
\end{eqnarray}
in the following condition $3 M u^2\ll u$ \cite{Hu}, where
\begin{equation}
u(\varphi)=\frac{M}{L^2}(1+e \cos\varphi)
\label{17}
\end{equation}
is the analytical elliptical solution which has already been found in Newton's gravity, and $e$ is the orbital
eccentricity which has been considered as a small constant.

Now on, we consider the HPM of solving equation (\ref{15}). For this end, we suppose the following homotopy
\begin{equation}\label{18}
u''+u-\frac{M}{L^2}-p\,\, 3 M u^2=0,\,\,\,\,\,\,p \in [0,1],
\end{equation}
where the prime denotes derivative with respect to $\varphi$, and assume that the solution of (\ref{15})
can be expressed as a series in $p$ by
\begin{equation}\label{19}
u(\varphi) = u_0(\varphi) + p\, u_1 (\varphi) + p^2 u_2(\varphi) + ...\, .
\end{equation}
According to (\ref{17}), the initial conditions for $u_0(0)$ and $u_i (0)$  can be chosen as follows
\begin{eqnarray}\label{20}
u_0(0)=\frac{M}{L^2}(1+e),~~~~u_0'(0)=0, \\
u_i(0)=u_i'(0)=0,\label{21}
\end{eqnarray}
where $i \geq 1$.
The substitution of (\ref{19}) into equation (\ref{18}) yields
\begin{eqnarray}
p^0&:&u_0''+u_0-\frac{M}{L^2}=0,   \label{22}\\
p^1&:&u_1''+u_1-3Mu_0^2=0,\label{23}\\
p^2&:&u_2''+u_2-6Mu_0 u_1=0, \label{24}\\
& &.\,\,.\,\,.\,\,.\,\,.\,\,.\,\,.\,\,.\,\,.\,\,.\,\,.\,\,. \nonumber
\end{eqnarray}
It is noteworthy that we obtain the set of linear equations. Their solutions with the initial conditions
(\ref{20}), (\ref{21}) can be readily found as
\begin{eqnarray}
u_0(\varphi)&=&\frac{M}{L^2}(1+e\cos \varphi), \nonumber\\
u_1(\varphi) &=& \frac{M^3}{L^4}\Big(3+2e^2+3e\varphi \sin\varphi\nonumber\\
&-&e^2 \cos^2\varphi-(3+e^2)\cos \varphi\Big),\label{25}
\end{eqnarray}
where we have deliberately limited our calculation by the minimum degree of approximation.  All subsequent
approximations can also be obtained easily.

In accordance with the HPM, it follows  from (\ref{19}) and (\ref{25}) that the solution of equation (\ref{15})
is given by
\begin{eqnarray}
u(\varphi)=\frac{M}{L^2}(1+e\cos \varphi)+\frac{M^3}{L^4}\Big(3+2e^2+3e\varphi \sin\varphi\nonumber\\
-e^2 \cos^2\varphi-(3+e^2)\cos \varphi\Big).\label{26}
\end{eqnarray}
Comparing our result (\ref{26}) with the approximate formula (\ref{16}), one can conclude that these solutions
are different in the term $ -(M^3/L^4)(3+e^2)\cos \varphi$,  and therefore they do not give the same value of
the perihelion shift.

In order to demonstrate the difference in the numerical and approximate solutions
of the equation (\ref{15}) more clearly, we take some hypothetical parameters of the system, providing significant
perihelion shift compared to that of Mercury. So we set $M=0.04, L=0.2$ in conventional units, and $e=0.8$.

\begin{figure}[thbp]
\centering
\includegraphics[width=0.5\textwidth]{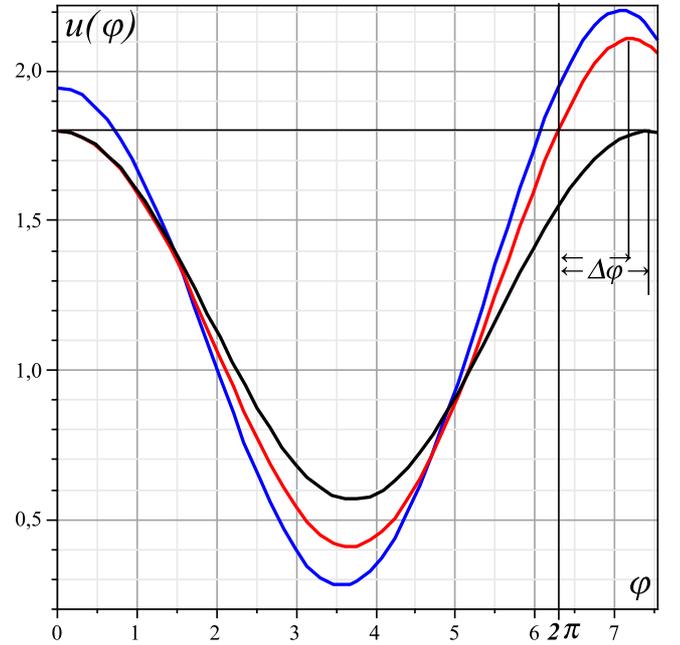}
\caption{The orbital motion $u(\varphi)$ is plotted for the numerical solution to (\ref{15}) (black line),
the approximation (\ref{16}) (blue line), and for the HPM solution (\ref{26}) (red line).}
\label{Figure_1}
\end{figure}

Typically, the perihelion shift is found from the appropriate approximate solutions, neglecting $e^2$ compared
to $e$. Thus, for small eccentricity, the equation (\ref{16}) can be rewritten as
$$
u(\varphi)=\frac{M}{L^2}[1+e \cos(\varphi-\epsilon \varphi)],
$$
where $\epsilon = 3M^2/L^2$. For the perihelion of orbit, it satisfies $cos(\varphi-\epsilon \varphi)=1$.
Therefore, the precession angle of perihelion equals to $\Delta \varphi = 2 \pi \epsilon = 6 \pi M^2/L^2$.
A more precise formula for the shift can be obtained from the condition $u'(\varphi) = 0$ in the perihelion,
and $\varphi = 2 \pi + \Delta \varphi $, assuming that $\Delta \varphi \ll 1 $. As applied to the approximate
solution (\ref{16}), this method yields
$$
\Delta \varphi = 6\pi \frac{M^2}{L^2}\Big[1-2(3+e)\frac{M^2}{L^2}\Big]^{-1}.
$$At
the same time, the use of this approach to our solution (\ref{26}) gives the following result
\begin{equation}\label{27}
\Delta \varphi_{HPM} = 6e\,\pi \frac{M^2}{L^2}\Big[e-(3+6e-e^2)\frac{M^2}{L^2}\Big]^{-1}.
\end{equation}
Note that even for a small value of eccentricity, we have $\Delta \varphi_{HPM} \ne\Delta \varphi$, but both
of them lead to the same expression for $\Delta \varphi \approx 6 \pi M^2/L^2$  when $L^2/M^2 \ll 1$.

Finally, let us obtain the HPM orbits and shift in the Reissner-Nordstorm spacetime of a charged star.
In this case, we have \cite{Weinberg}
\begin{equation}
f(u)=h(u)=1-2M u + Q^2 u^2, \label{28}
\end{equation}
where $Q$ is the charge. According to (\ref{8}) and (\ref{28}), the main equation (\ref{15}) is replaced
by the following one
\begin{equation}
\frac{d^2 u}{d\varphi^2}+\Big(1+\frac{Q^2}{L^2}\Big)u=\frac{M}{L^2}+3\,u^2\,M-2Q^2u^3.
\label{29}
\end{equation}
Assuming that the unperturbed equation should have solution (\ref{8}), consider the following homotopy
\begin{equation}\label{30}
u''+u-\frac{M}{L^2}+p\,\Big(\frac{Q^2}{L^2}u-3 M u^2+2Q^2u^3\Big)=0,
\end{equation}
where  $p \in [0,1]$. Substituting (\ref{19}) into equation (\ref{30}), we get
\begin{eqnarray}
p^0&:&u_0''+u_0-\frac{M}{L^2}=0,   \label{31}\\
p^1&:&u_1''+u_1+\frac{Q^2}{L^2}u_0-3Mu_0^2+2Q^2u_0^3=0,\label{32}\\
& &.\,\,.\,\,.\,\,.\,\,.\,\,.\,\,.\,\,.\,\,.\,\,.\,\,.\,\,. \nonumber
\end{eqnarray}
where  the simplest approximation is taken. The set of linear equations (\ref{31}), (\ref{32}) with the
initial conditions (\ref{20}), (\ref{21}) can be easily solved, giving
\begin{eqnarray}
u&=&\frac{M}{L^2}(1+e\cos \varphi)+\frac{M^3}{L^4}\Big[3+2e^2
-2(1+2e^2)\frac{Q^2}{L^2}\nonumber\\&-&\frac{Q^2}{M^2}+\Big(1-\frac{Q^2}{L^2}+\frac{Q^2}{6M^2}
-e^2\frac{Q^2}{4L^2}\Big)3e\varphi \sin\varphi \nonumber\\
&-&\Big(3+e^2+(e^3-8e^2-8)\frac{Q^2}{4L^2}-\frac{Q^2}{M^2}\Big)\cos \varphi \nonumber\\
&-&\Big(1-2\frac{Q^2}{L^2}\Big)e^2 \cos^2\varphi+\frac{Q^2}{4L^2}e^3 \cos^3 \varphi\Big].\label{33}
\end{eqnarray}
for the approximate solution $u(\varphi)=u_0(\varphi)+u_1(\varphi)$. With the help of our
solution (\ref{33}), one can easily obtain the value of shift angle. We do not provide it here
due to a cumbersome nature of its expression.

\section{\Large{Conclusions}}
\quad

Thus, in this work  we have considered a simple analytical computation of
the perihelion precession in General Relativity with the help of the  Homotopy Perturbation
Method. First, we have studied the example of geodesic motion in the Schwarzschild metric,
in order to approbate  HPM in the problem of planetary motion in GR, and present the main
steps in solving by this method. The comparison of our solution (\ref{26}) with the approximate
solution (\ref{16}), obtained earlier by the perturbation method, demonstrates the better degree
of accuracy. This result has been shown in Fig 1.

A direct result of the differences in solutions
consists in  that the perihelion precession (\ref{27}),  derived from the HPM solution (\ref{26}), yields
the better accuracy compared to the previously known result. It is important that
formula (\ref{27}) for $\Delta \varphi_{HPM}$ differs  from the standard  $\Delta \varphi$, even for the
small values of eccentricity. Only in the case of small parameter $L^2/M^2 \ll 1$,
the HPM approximation could give the commonly used value of precession  $6 \pi M^2/L^2$.
It is worthy to note, that all these results were obtained  by the single iteration.
This may give us the hope that the next iteration could provide us   the result with much
greater accuracy and a minimum size of computations.

Moreover, in our work we have obtained HPM solution for the Reissner-Nordstorm spacetime of a charged star.
For this case, the additional terms in the HPM approximation were obtained as compared to the similar solution
in the standard approximation (see, e.g. \cite{Hu}).

According to the results of this work and our work published earlier \cite{Shchigolev1}, we could make the following remarks.
Foremost, the undoubted advantage of this method consists of that there is no need to establish a small parameter for
solving a problem in some approximation, because such a small parameter sometimes could destroy the main feature of the
exact solution. On the other hand,  the approximate solution,  and the rate of its convergence in this method
greatly depend on the construction of homotopies.
Nevertheless, we argue that HPM is able to provide a rather high degree of accuracy in the problems of
astrophysics and cosmology associated with nonlinearity of their main equations.

\noindent\hrulefill


\begin{thebibliography}{99}
{\small
\bibitem{Weinberg}  S. Weinberg.  Gravitation and Cosmology: Principles and Applications of The General Theory of Relativity,
John Wiley. Press, New York, 1972.

\bibitem{Magnan} C. Magnan. Complete calculations of the perihelion precession of Mercury and the deflection of light by
the Sun in General Relativity, arXiv:0712.3709


\bibitem{Hu} Ya-Peng Hu, Hongsheng Zhang, Jun-Peng Hou, and Liang-Zun Tang. Perihelion precession and deflection of light
in the general spherically symmetric spacetime, arxiv: 1312.7419.

\bibitem{Ara} H. Arakida. Note on the Perihelion/Periastron Advance Due to Cosmological Constant, Int. J. Theor.
Phys. 52, 1408–1414, 2013; DOI 10.1007/s10773-012-1458-2.


\bibitem{Wells} A. A. Vankov. General Relativity Problem of Mercury's Perihelion Advance Revisited,  arXiv:1008.1811.


\bibitem{Pejic}  M. Pejic.  Calculating perihelion precession using the multiple scales method,
http://math.berkeley.edu/ ~mpejic/pdfdocuments/PerihelionPrecession.pdf

\bibitem{Cuzinatto} R. R. Cuzinatto, P. J. Pompeia, M. de Montigny, F. C. Khanna, J. M. Hoff da Silva.
Classic tests of General Relativity described by brane-based spherically symmetric solutions,
Eur. Phys. J. C  74, 3017, 2014; DOI 10.1140/epjc/s10052-014-3017-x.

\bibitem{Fokas} A.S. Fokas, C.G. Vayenas, D. Grigoriou. Analytical computation of the Mercury perihelion
precession via the relativistic gravitational law and comparison with general relativity,  arXiv:1509.03326.

\bibitem{Ruggiero} M. L. Ruggiero. Perturbations of Keplerian orbits in stationary spherically symmetric
spacetimes, Int. J. Mod. Phys. D, Vol. 23, No. 5, 1450049, 2014.

\bibitem{Saridakis} L. Iorio, E. N. Saridakis.  Solar system constraints on $f(T)$ gravity, Mon. Not. Roy.
Astron. Soc., 427,  1555, 2012; DOI:  10.1111/j.1365-2966.2012.21995.x

\bibitem{Kraniotis} G. V. Kraniotis, S. B. Whitehouse. Compact calculation of the Perihelion Precession of
Mercury in General Relativity, the Cosmological Constant and Jacobi's Inversion problem, Class. Quant. Grav.
20, 4817-4835, 2003; DOI:  10.1088/0264-9381/20/22/007

\bibitem{He} J.-H. He. Homotopy perturbation technique, Comput. Meth. Appl. Mech. Eng., Vol. 178, 257-262, 1999.

\bibitem{He2} J.-H. He. A coupling method of homotopy technique and perturbation technique for nonlinear problems,
Int. J. Nonlinear Mech., 35, No. 1, 37–43, 2000.

\bibitem{He3} J.-H. He. Homotopy perturbation method: a new nonlinear analytical technique, Appl. Math. Comput.,
Vol. 135, 73–79, 2003.

\bibitem{He4} J.-H. He. Application of Homotopy Perturbation Method to Nonlinear Wave Equations,
Chaos Solitons $\&$ Fractals, 26, 695-700, 2005.

\bibitem{He6} J.-H. He. Homotopy perturbation method with two expanding parameters,
Indian J. Phys.; DOI 10.1007/s12648-013-0378-1

\bibitem{Cveticanin} L. Cveticanin. Homotopy-perturbation method for pure nonlinear
differential equation, Chaos, Solitons $\&$ Fractals, vol. 30, No. 5, 1221 - 1230, 2006.

\bibitem{Nourazar} S.S. Nourazar,  M. Soori, A. Nazari-Golshan.   On the Exact Solution of
Newell-Whitehead-Segel Equation Using the Homotopy Perturbation Method,
Australian Journal of Basic and Applied Sciences, 5, 1400-1411, 2011.


\bibitem{He7} J.-H. He. Asymptotic Methods for Solitary Solutions
and Compactons, Abstract and Applied Analysis,
Volume 2012, Article ID 916793, 130 pages
doi:10.1155/2012/916793

\bibitem{Shchigolev1} V.  Shchigolev. Homotopy Perturbation Method for Solving a Spatially
Flat FRW Cosmological Model, Universal Journal of Applied Mathematics 2(2): 99-103, 2014 DOI: 10.13189/ujam.2014.020204

\bibitem{Rahaman}F. Rahaman, S. Ray, A. Aziz, S. R. Chowdhury, D. Deb. Exact Radiation Model
For Perfect Fluid Under Maximum Entropy Principle,  arXiv:1504.05838.

}

\end{thebibliography}
\end{document}